\UseRawInputEncoding
\documentclass[aps,reprint,amsmath,amssymb]{revtex4-1}
\usepackage{graphicx,subfigure}
\usepackage{epstopdf}
\usepackage{upgreek}
\usepackage{bm}

\usepackage{dcolumn}
\usepackage{epstopdf}
\usepackage{color}

\makeatletter
\def\btt#1{\texttt{\@backslashchar#1}}%
\DeclareRobustCommand\bblash{\btt{\@backslashchar}}%
\makeatother

\begin{document}

\title{Critical splay fluctuations and colossal flexoelectric effect above the nonpolar to polar nematic phase transition}

\date{\today}
\author{Abinash Barthakur$^{1}$, Jakub Karcz$^{2}$, Przemyslaw
 Kula$^{2}$ and Surajit Dhara}
\email{surajit@uohyd.ac.in} 
\affiliation{$^{1}$School of Physics, University of Hyderabad, Hyderabad-500046, India\\
$^{2}$Institute of Chemistry, Faculty of Advanced Technologies and Chemistry, Military University of Technology, Warsaw, Poland  
}

\begin{abstract}
 The recent discovery of splay nematic liquid crystals with polar order (ferroelectric) has created immense interest. Despite intensive research, several physical properties of this liquid crystal are yet to be investigated and understood. Here, we report experimental studies on the surface alignment, birefringence and flexoelectric coefficient of a polar nematic (N\textsubscript{F}) liquid crystal. Our experiments directly reveal that the splay fluctuations influence the birefringence several degrees above the nonpolar (N) to polar-nematic (N\textsubscript{F}) phase transition temperature and the heat capacity exponent obtained from the tilt angle fluctuations is close to the value reported in adiabatic scanning calorimetry measurements. The flexoelectric coefficient of the nonpolar nematic phase shows a power-law dependence on temperature. Our results demonstrate a strong coupling of splay fluctuations with electric polarisation that reduces the splay elastic constant and consequently enhances the flexoelectric coefficient. These results are important for forthcoming applications as well as for understanding all pretransitional effects in ferroelectric nematic liquid crystals.
\end{abstract}
\preprint{HEP/123-qed}
\maketitle

%\section{Introduction}
Ordinary nematic liquid crystals (LCs) made of elongated molecules have only long-range orientational order. The average orientation direction of the long axes of the molecules is called the director and denoted by a unit vector $\hat{n}$ that possesses inversion symmetry~\cite{pg} i.e, $\hat{n}\equiv-\hat{n}$. The inversion symmetry in the director field is the key for the lack of local macroscopic polarisation even though the constituting molecules have non-zero net dipole moments. In liquid crystal devices (LCDs) mostly the nonpolar liquid crystals are utilised in which the reorientation of the director and resulting electrooptic effects are due to the external electric field~\cite{blinov}. Here, the bulk nematic responds due to the dielectric quadrupolar coupling. Recently, a novel nematic phase, known as splay nematic phase (N\textsubscript{S}) has been discovered in compounds with highly polar and wedge-shaped molecules~\cite{RJM1,RJM2,NH,MA,AM}. This phase is composed of periodic splay deformation with polar domains oriented along the director and identified as the ferroelectric nematic phase (N\textsubscript{F})~\cite{XC,NS}. The spontaneous polarisation \textbf{P} facilitates a linear coupling of the director $\hat{n}$ with the external electric field \textbf{E}, which is lacking in nonpolar nematic liquid crystals. 

In spite of intensive research, many physical properties of this new variant of nematic liquid crystal are still unexplored~\cite{REV}. The physical, optical, as well as electro-optical properties of the ferroelectric nematic LCs, reported so far turn out to be intriguing and promising for new applications~\cite{PR,SB,HN1,OD,CLF,JLI,XZ}. However, application as well as physical measurements require defect-free uniform alignment of the nematic director between the confining substrates. Detailed surface alignment studies of ferroelectric nematic LC (RM-734) show that homeotropic alignment of the ferroelectric nematic is still a matter of concern~\cite{FC,NS,MA}. Further, precession birefringence measurements across the N to N\textsubscript{F} phase transition and the flexoelectric responses have not been reported. In this letter, we report experimental studies on compound RM-734 that exhibits nonpolar nematic (N) to ferroelectric nematic (N\textsubscript{F}) phase transition. We show that  JALS-204 provides a good homeotropic alignment of the N phase.  We measure the splay fluctuations directly from the temperature-dependent birefringence and obtain a critical exponent related to the heat capacity. Furthermore, we measure the flexoelectric coefficient of the nonpolar nematic phase that shows a considerable increase as N to N\textsubscript{F} transition temperature is approached.
%#####################
\begin{figure}
\centering
\includegraphics[scale=0.36]{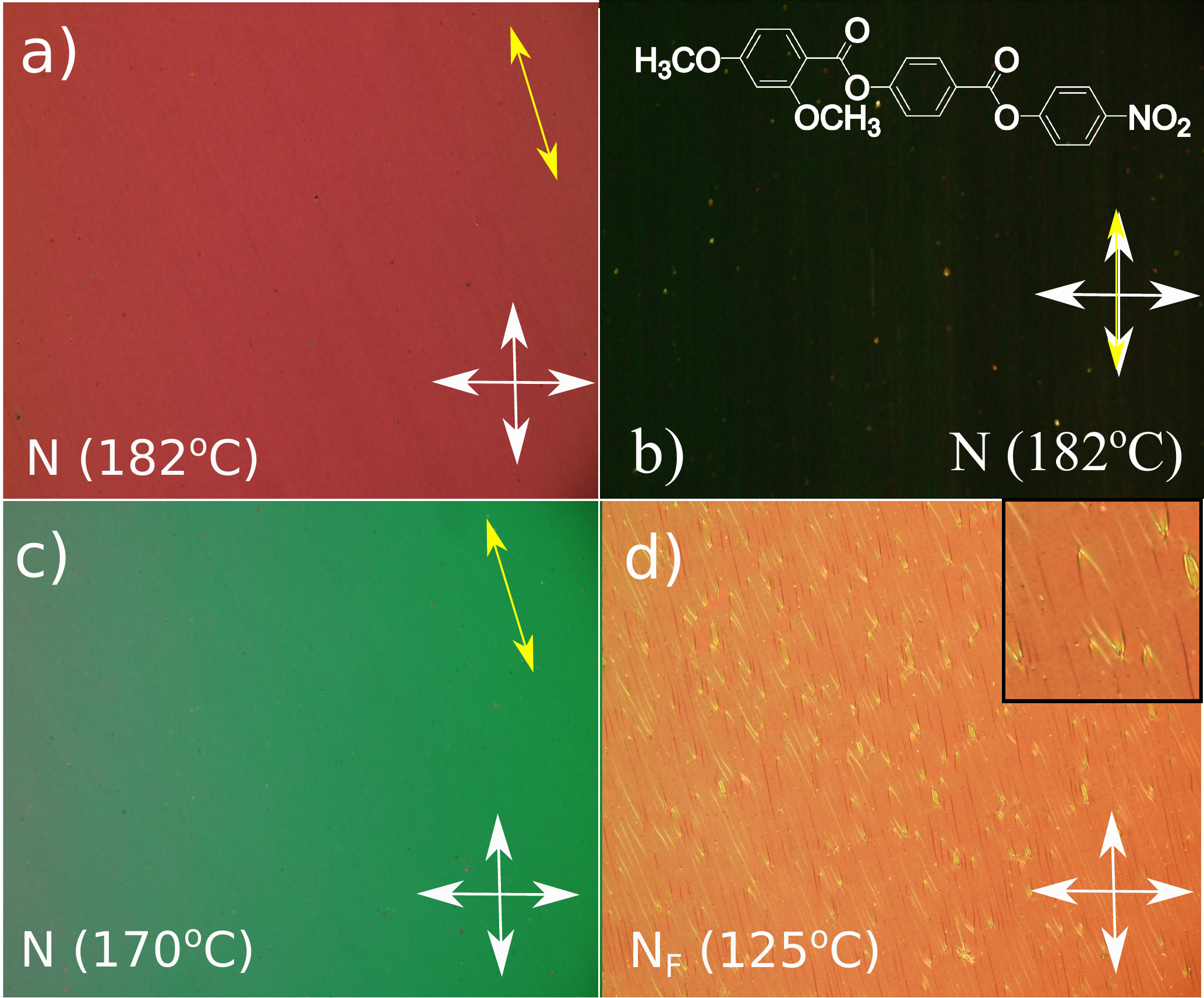}
\caption{Optical polarising microscope textures observed in PI-2555 coated cells. Cell thickness $d=8.1~\mu$m. (a) Sample filled in the N phase. (b) Texture observed under crossed polarisers. The molecular structure is superimposed on the texture. (c) The sample was heated to the isotropic phase and cooled down to the N phase. (d) Cooled down to the N\textsubscript{F} phase from the N phase. The box at the corner shows an enlarged view of lens-shaped domains. The width of the images is 1.3 mm.}
\label{fig:figure1} 
\end{figure}
%#####################

%\section{Experimental}
The studied compound RM-734 was synthesized in our laboratory following our new method, enhancing the overall yield~\cite{PK}.
 The molecular structure is shown in Fig.\ref{fig:figure1}(b). On cooling, the sample shows the following phase transitions under optical polarising microscope:  I 185\textsuperscript{o}C N 130\textsuperscript{o}C N\textsubscript{F}. We used indium-tin-oxide (ITO) coated glass plates for making cells for the optical birefringence measurements. We used PI-2555 (HD MicroSystems) for homogeneous alignment and JALS-204 (JSR Corporation) and SE-1211 (Nissan Chemicals) for the homeotropic alignment of the sample. These polyimides were spin-coated on glass substrates and cured at an appropriate elevated temperature and rubbed in an antiparallel way.  
In all experiments, the LC sample was filled at the high-temperature nematic phase. It facilitates a flow-induced homogeneous alignment of the molecules along the rubbing direction.   
Measurement of birefringence was carried out using a phase modulation technique~\cite{BI} and the data were collected at temperature steps of 100 mK. The setup includes a photoelastic modulator (PEM100, Hinds Instruments), a He-Ne laser, a lock-in amplifier (Amtek 7265) and an Instec temperature controller (MK2000). This setup is capable of giving optical retardation measurement sensitivity down to 0.1 nm~\cite{ACC}. Flexoelectric measurements were carried out in hybrid-aligned nematic (HAN) cells that stabilise splay-bend texture due to antagonistic homogeneous (PI-2555) and homeotropic (JALS-204) anchoring of the two substrates~\cite{Flexo}.  
 The maximum twist angle is given by $\phi = \frac{e^{*}d}{\pi K}E$, where $e^{*}=e_s-e_b$ is the flexoelectric coefficient, $K$ is the average elastic constant and $d$ is the cell thickness~\cite{Flexo}. To measure $\phi$ the analyzer was rotated across the extinction state and the angle-dependent intensity was measured by a photomultiplier tube (Hamamatsu, H6780-01). The slope of the twist angle ($\phi$) versus field $E$ across the zero-field provides $e^{*}/K$ for a known cell gap.

%#####################
\begin{figure}
\centering
\includegraphics[scale=0.36]{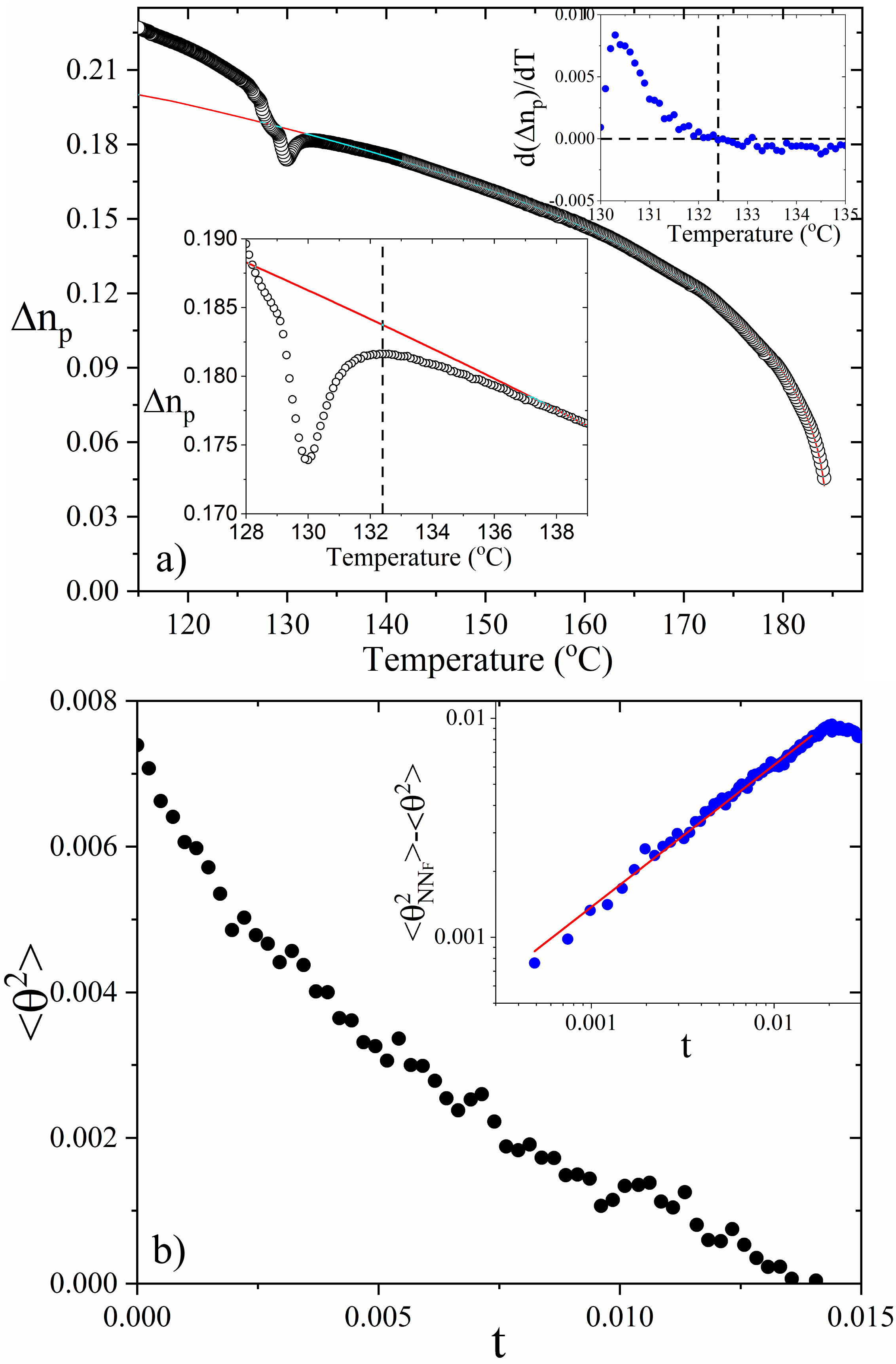}
\caption{(a) Variation of birefringence ($\Delta n_p$) in a planar or homogeneous cell with temperature. Red curve shows the Haller extrapolation with fit parameters $\beta=0.3$ obtained by fitting the data away from the pre-transitional suppression region. The bottom inset shows the enlarged view of reduction in birefringence across the N-N\textsubscript{F} phase transition temperature. The vertical line indicates the onset temperature of N-N\textsubscript{F} coexistence region.  The top inset shows the temperature derivative of $\Delta n_p$ across the phase transition. (b) Variation of the mean square fluctuations $\langle\theta^2(T)\rangle$ of the tilt angle calculated using Eq.(2) as a function of reduced temperature $t=\frac{T-T_{NN_F}}{T_{NN_F}}$. Inset shows the relative tilt fluctuations $\langle\theta_{NN_F}^{2}\rangle-\langle\theta^2(T)\rangle\sim t^{1-\alpha}$  and the red line shows a power-law fit to the date with the exponent $\alpha=0.36\pm0.01$. Cell thickness $d=11.9~\mu$m.  }
\label{fig:figure2} 
\end{figure}
%##################### 

%\section{Results and Discussion}

 First, we checked the quality of the homogeneous alignment of the sample in PI-2555 coated cells by optical polarising microscopy. We filled the sample in the N phase (Fig.\ref{fig:figure1}(a)) then heated it to the isotropic phase and cooled it back to the N phase (Fig.\ref{fig:figure1}(c)). The homogeneous alignment is recovered in the N phase. The uniform alignment is also evident from the cross-polarised image with $\hat{n}$ parallel to the polariser (Fig.\ref{fig:figure1}(b)). Then the sample was cooled into the N\textsubscript{F} phase (Fig.\ref{fig:figure1}(d)) in which some lens-shaped domains elongated along the rubbing direction were observed. These domains are similar to those reported by Chen \textit{et al.}~\cite{XC} where the domain boundary separates two regions with the opposite orientation of in-plane polarisation \textbf{P}. Overall, the homogeneous alignment of the N phase in PI-2555 coated cells is very good and reproducible under the heating and cooling cycles across the nematic to isotropic (N-I) phase transition. 
  
  We measured the temperature-dependent birefringence ($\Delta n_p$) of the sample while cooling. As shown in Fig.\ref{fig:figure2}(a) the birefringence increases with decreasing temperature as expected followed by a dip. The sample shows a very weak first-order  N to N\textsubscript{F} phase transition~\cite{JT}. Hence, the dip is presumably due to the coexistence of polar and nonpolar domains and defects at the transition region. The high-temperature birefringence data can be fitted to the Haller formula~\cite{IH}: 
   \begin{equation}
  \Delta n_p=\Delta n_0 (1-T/T^{*})^{\beta}
   \end{equation}
   where $\Delta n_{0}$ is the birefringence of the perfectly aligned nematic phase with orientational order parameter $S=1$ and $\beta$ is the critical exponent connected to the N-I phase transition temperature. Careful observation shows that there is a  change of curvature around 137$^\circ$C (see inset of Fig.\ref{fig:figure2}(a)). As observed $\Delta n_p$ data fits well above 137$^\circ$C and below this temperature, a clear departure (reduction) from the predicted path is observed. In the N\textsubscript{F} phase the director is splayed periodically. Hence, the reduction of birefringence above the N-N\textsubscript{F} phase transition temperature is a clear indication of strong splay fluctuations of the director which increases as the N-N\textsubscript{F} transition temperature is approached.  Such fluctuations lead to the tilting of the molecules from the mean direction, resulting in reduced birefringence. The onset temperature of the fluctuations nearly coincides with the temperature where the splay elastic constant measured from the dynamic light scattering experiments begins to decrease drastically~\cite{MA}. We make a quantitative estimation of the tilt fluctuations from the temperature-dependent birefringence above the N-N\textsubscript{F} phase transition temperature. The mean square fluctuation tilt angle $\langle\theta^2\rangle$ can be obtained  from the decrease of birefringence using following equation~\cite{MS,JF}:
 \begin{equation}
  \Delta n_p(T)=\Delta n_0 (1-\frac{3}{2}\langle\theta^2\rangle)
 \end{equation}	
  where  $\Delta n_p$ is the measured value and $\Delta n_0$ is obtained from the extrapolated birefringence (see Eq.(1)) in the absence of fluctuations.  Figure \ref{fig:figure2}(b) shows the calculated $\langle\theta^2\rangle$ with reduced temperature $t=\frac{T-T_{NN_F}}{T_{NN_F}}$, where $T_{NN_F}$ is the N to N\textsubscript{F} phase transition temperature. The macroscopic tilt and tilt fluctuation angle near the phase transition can not be separated directly. Hence, we find $T_{NN_F}(=132.3^\circ$C) from the temperature derivative of the birefringence at which the slope changes sign as shown in the inset of Fig.\ref{fig:figure2}(a). The fluctuations are large near the N-N\textsubscript{F} phase transition temperature and it decreases rapidly and finally goes to zero a few degrees above the transition (Fig. \ref{fig:figure2}(b)). Theoretically, the critical component of the fluctuations is known to vary with reduced temperature as~\cite{KCL,SER,DP}: 
  \begin{equation}
  \langle\theta_{NN_F}^{2}\rangle- \langle\theta^2\rangle\sim t^{1-\alpha}, \hspace{0.5cm} \text{for} \hspace{0.5cm} t>0
  \end{equation}
   where $\alpha$ is the exponent of the heat capacity. The inset of Fig. \ref{fig:figure2}(b) shows the variation of tilt fluctuations with reduced temperature in the interval of $t=5\times10^{-4}$ to $t=1.5\times10^{-2}$. The exponent obtained from the linear fit is found to be $\alpha=0.36\pm0.01$.   This is reasonably closer to the heat capacity exponent $\alpha=0.5$ reported in recent adiabatic scanning calorimetry measurements~\cite{JT}. The experimental values of the exponents $\alpha=0.36$ and $\beta=0.3$ indicates the phase transition is a tricritical type located at the crossover from first order to second order along a phase transition line. It should be mentioned that the good quality of the alignment of the sample is very important for the observation of the effect of fluctuations on temperature-dependent birefringence.\\
   % It may be mentioned that similar fluctuations leading to instantaneous heliconical states have been reported above the nematic to nematic twist-bend (N\textsubscript{TB}) phase transition temperature~\cite{DP}.
  
   %#####################
\begin{figure}[t]
\centering
\includegraphics[scale=0.36]{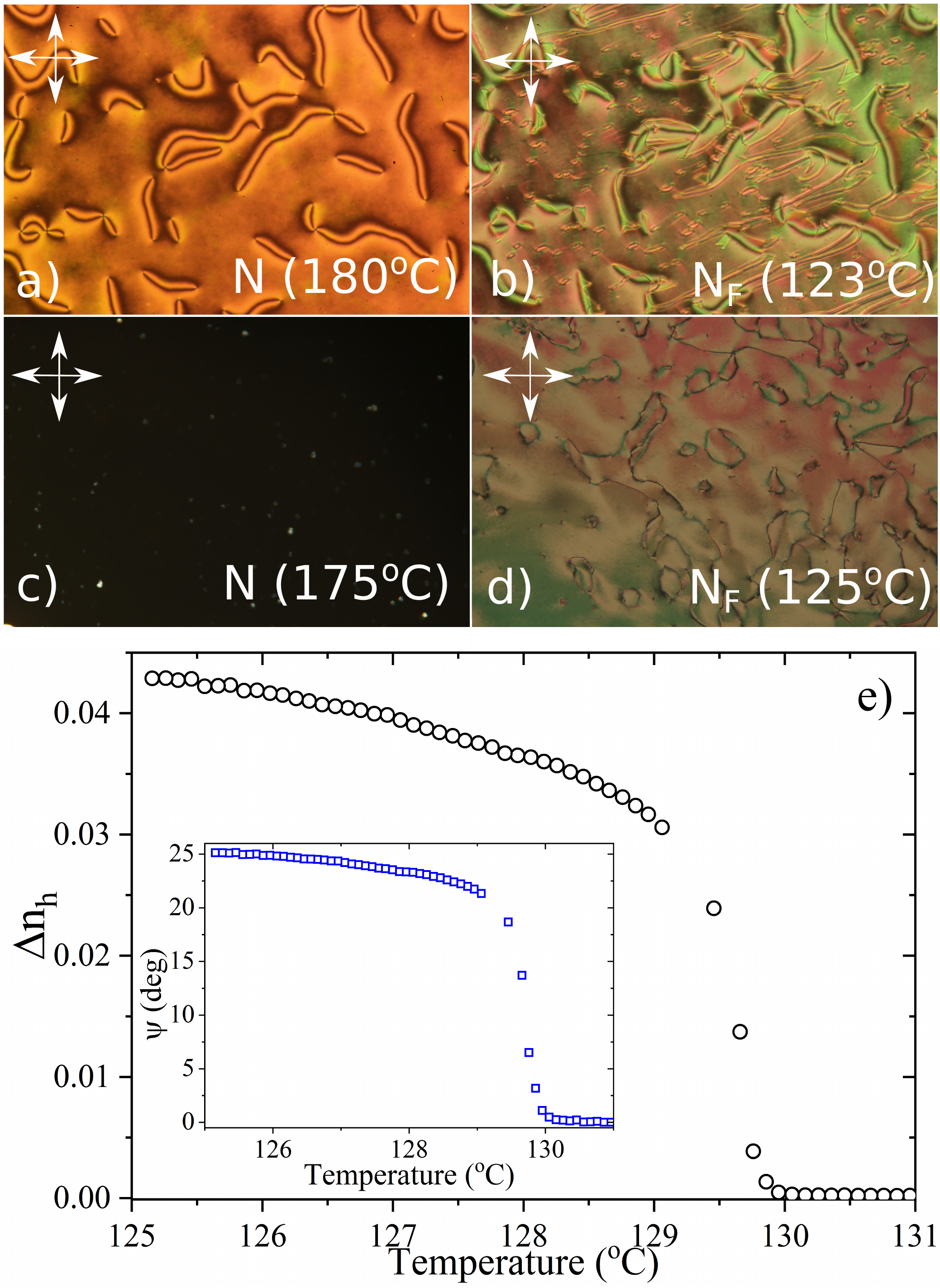}
\caption{Optical polarising microscope textures of the (a) N and (b) N\textsubscript{F} phases in SE-1211 coated cell. Textures of the (a) N and (b) N\textsubscript{F} phases in JALS-204 coated cell. (e) Temperature-dependent birefringence ($\Delta n_h$) in a JALS-204 cell across the N-N\textsubscript{F} transition. Inset shows the estimated tilt angle $\psi$ as a function of temperature. Cel thickness $d=8.2~\mu$m for JALS-204 and $d=8.7~\mu$m for SE-1211 coated cells. The width of the images is 1.3 mm.
\label{fig:figure3}}
\end{figure}
%#####################

 While the homogeneous alignment of RM-734 was found to be rather successful the homeotropic alignment of the sample has been found to be unsuccessful~\cite{MA,NS,REV}. It has been reported that both the nematic phases of RM-734 can not be aligned perfectly homeotropically~\cite{MA}. In hydroxylated glass cells, the N phase exhibited uniform homeotropic alignment but an inhomogeneous planar-type texture was observed in the N\textsubscript{F}~\cite{FC}. In order to address this issue we experimented with two widely used homeotropic alignment layers, namely polyimides SE-1211 and JALS-204. The sample in both the cells was filled in the isotropic phase and cooled at the rate of 0.5$^\circ$C/min to the nematic phase. The textures in SE-1211-coated cells in both phases appear to be Schlieren type and not homeotropic (Fig.\ref{fig:figure3}(a,b)). In JALS-204 coated cells, the N phase is homeotropic (uniform dark state) and a low-birefringent texture is observed below the N-N\textsubscript{F} transition temperature. It suggests that the director is tilted on the substrate in the N\textsubscript{F} phase. To get a quantitative estimate of the tilt angle we measured the temperature-dependent birefringence ($\Delta n_h$) in a homeotropic cell. The birefringence increases from zero as expected and tends to saturate around $\Delta n_h=0.043$ which is much smaller compared to the birefringence $\Delta n_p\simeq 0.25$ of the homogeneous cell (see Fig.\ref{fig:figure2}(a)). It reassures that the director in the N\textsubscript{F} phase is tilted at an angle with respect to the substrate normal. Assuming the director tilts by an angle $\psi$ with respect to the substrate normal (i.e., the propagation direction of light) the effective refractive index of the nematic LC can be written as $1/n^2_{eff}=\cos^2(\psi)/n_o^2+\sin^2 (\psi)/n_e^2$, where $n_e$ and $n_o$ are the extraordinary and ordinary refractive indices. The effective  birefringence of the homeotropic cell can be approximately expressed as:
 \begin{equation}
 \Delta n_h=n_{eff}-n_o=\sin{^2}{\psi}\left (\Delta n_p-\frac{2(\Delta n_p)^2}{n_o}\right	)
 \end{equation}
 Assuming $\Delta n_p <<n_o$, we can estimate the tilt angle $\psi$ from the following equation:
 \begin{equation}
  \psi=\sin^{-1}\sqrt{\Delta n_h/\Delta n_p}
 \end{equation}
    The inset of Fig. \ref{fig:figure3}(e) shows the calculated tilt angle with temperature. It almost jumps to 20$^\circ$ from zero and saturates to about 25$^\circ$. 
      
%%%%%%%%%%%%%%%%%%%%%%%%
\begin{figure}[t]
\centering
\includegraphics[scale=0.42]{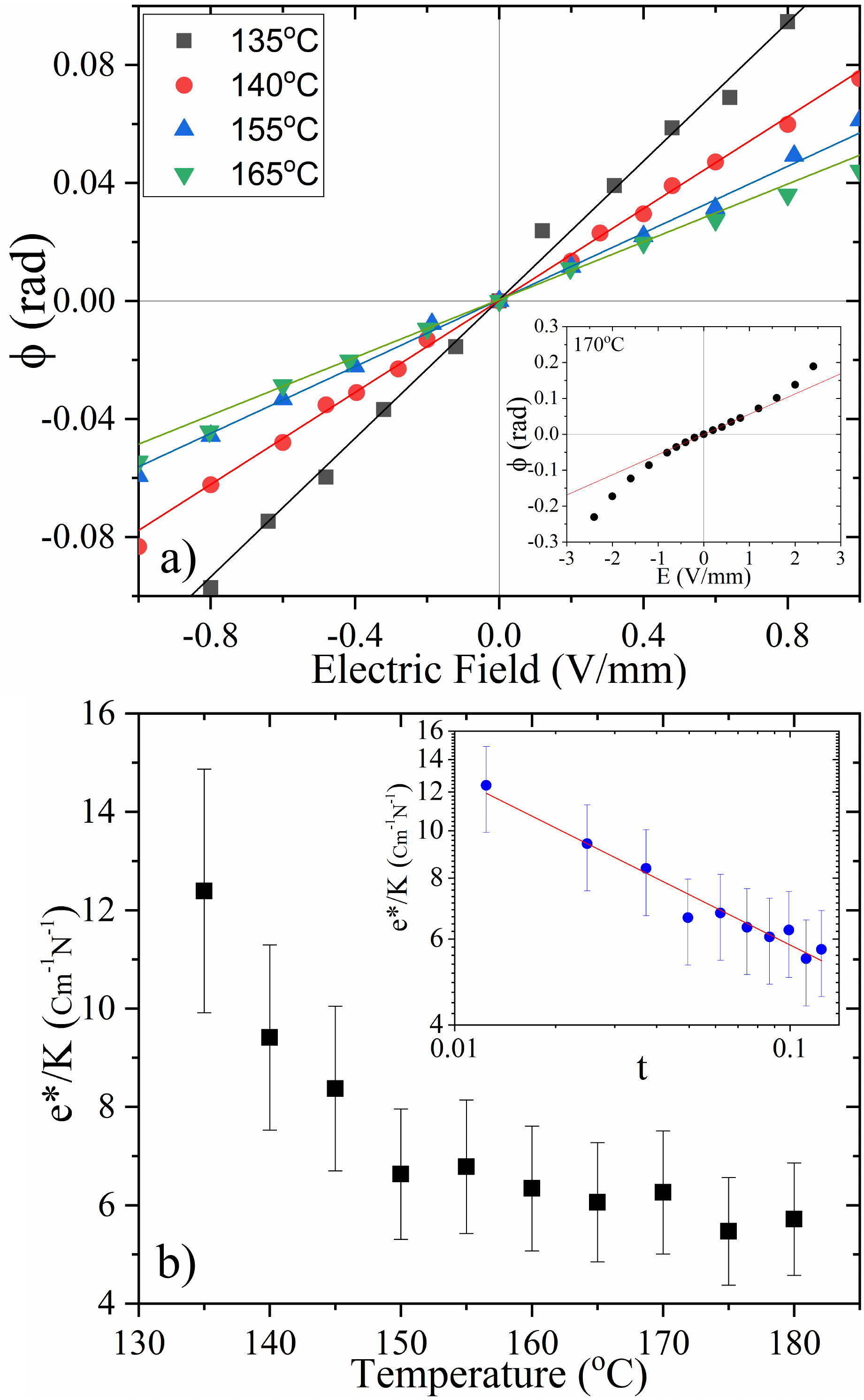}
\caption{(a) Variation of maximum twist angle $\phi$ with dc electric field at different temperatures showing linear coupling due to flexoelectric polarisation. Solid lines are linear fits to the data. Inset shows the variation of $\phi$ at 170\textsuperscript{o}C showing takeover of dielectric behaviour at higher field. (b) Variation of $e^{*}/K$ obtained from the slope at different temperatures. Error bars represent the standard deviation of the mean. Inset shows the power-law fit to: $e^{*}/K\sim t^{-0.34\pm{0.01}}$. Cell thickness $d=25.2~\mu$m.  
\label{fig:figure4}}
\end{figure} 
%%%%%%%%%%%%%%%%%%%%%%%%

In what follows, we measured the flexoelectric coefficient of the sample of the nonpolar nematic phase. The molecules are wedge-shaped and have a large axial electric dipole moment~\cite{MA} $\mu_z\simeq11.3$D hence expected to be capable of demonstrating a larger flexoelectric response than ordinary nematic LCs with smaller dipole moments and cylindrical symmetry. Nevertheless, so far no flexoelectric measurements on RM-734 have been reported. We measured the flexoelectric response in a hybrid-aligned cell following the standard protocol as described in the experimental section. The inset of Fig. \ref{fig:figure4}(a) shows the variation of maximum twist angle $\phi$ with dc electric field at 170$^\circ$C. In the low field range ($\leq 1.0$ V/mm) the variation is linear due to the coupling of the flexoelectric polarisation with the applied field and the slope change at the higher field is due to the dielectric coupling. To avoid repeated high field exposure and any permanent distortion which may arise due to high field, we restricted our measurements within the low field range ($\leq 1.0$ V/mm), where $\phi$ varies linearly with the field as shown in Fig.\ref{fig:figure4}(a). The positive slope of $\phi$ at all temperatures shows the flexoelectric coefficient of RM-734 is positive. The coefficient $e^{*}/K$ obtained at different temperatures above the N-N\textsubscript{F} transition temperature is shown in Fig.\ref{fig:figure4}(b). Here, two important observations are made. Firstly, the coefficient $e^{*}/K$ is nearly two to three times larger than ordinary low molecular weight cyanobiphenyl compounds having smaller dipole moments~\cite{FL1,FL2}.  Secondly, as the temperature is decreased and the system approaches the N-N\textsubscript{F} transition temperature, $e^{*}/K$ increases considerably. In fact, the temperature dependence can be fitted to a power-law: $e^{*}/K\sim t^{-0.34\pm{0.01}}$  (see inset of Fig.\ref{fig:figure4}(b)). For dipolar contributions to the flexoelectricity~\cite{Flexo},  $e^{*}\propto S^2$ and according to the mean-field approximation~\cite{pg}, $K\propto S^2$, hence $e^{*}/K$ is expected to be independent of temperature as reported in many compounds.  Thus, the power-law dependence of $e^{*}/K$ of compound RM-734 is rather unusual and it can be explained based on two possible contributions. Firstly, the splay elastic constant ($K_{11}$) decreases more rapidly than predicted by the mean field as the N-N\textsubscript{F} transition temperature is approached. Secondly, there could be polar domains due to spontaneous splay fluctuations as the temperature is decreased which could enhance the flexoelectric coefficient. The effects of such domains have been observed recently in electroviscous studies~\cite{PK}. 
 
% \section{Conclusion}
In conclusion, we showed that the birefringence exhibits critical behaviour due to strong splay fluctuations several degrees above the N-N\textsubscript{F} phase transition temperature. The estimated specific heat exponent $\alpha$ obtained from the tilt angle fluctuations is close to the value reported in recent adiabatic scanning calorimetry measurements. Polyimide JALS-204 provides a perfect homeotropic and tilted director orientation in the N and N\textsubscript{F} phases, respectively. In sharp contrast to the ordinary nematic LCs, the flexoelectric coefficient of RM-734 shows a power-law dependence on temperature. It is accounted for by the effect of spontaneous splay fluctuations that reduces the splay elastic constant significantly and growing polar domains. A microscopic theory considering both contributions is required to understand the temperature dependence of the flexoelectric coefficient. The colossal flexoelectric effect is responsible for driving the ferroelastic to ferroelectric nematic transition via flexoelectric coupling as reported by Sebasti\'{a}n \textit{et al.}~\cite{NS}. Our results are important for all electrooptical applications of ferroelectric nematic liquid crystals, particularly for the liquid crystal display devices that exploit flexoelectric effect~\cite{CV,HJC}.\\
       
\noindent\textbf{Acknowledgments}: SD acknowledges financial support from SERB (Ref. No:CRG/2019/000425). AB acknowledges UGC-CSIR for fellowship. This work was supported by the National Science Centre grant 2019/33/B/ST5/02658. We acknowledge Arun Roy for his useful discussions. \\
\textbf{Conflict of Interest:}
The authors have no conflicts to disclose.\\
\textbf{Author Contributions:}
Surajit Dhara (Conceptualisation, data analysis and writing), Abinash Barthakur (Experiments, data analysis and writing), Jakub Karcz and Przemyslaw Kula (Synthesis).\\ 
\noindent\textbf{DATA AVAILABILITY:}
The data that support the findings of this study are available
within the article.\\\\


\begin{thebibliography}{99}


\bibitem{pg} de Gennes, P. G. The Physics of Liquid Crystals; Oxford University Press: Oxford, England, (1974).

\bibitem{blinov} L. M. Blinov, and V. G. Chigrinov, \textit{Electrooptic effects in liquid crystal materials}, (Springer, New York, 1994).

%\bibitem{MB} M. Born Sitzungsber Preuss Akad Wiss. \textbf{30}, 614-650 (1916). 

\bibitem{RJM1}	R. J. Mandle, S. J. Cowling, J. W. Goodby,  Chem. A Eur. J.\textbf{23}, 14554 (2017).

\bibitem{RJM2}	R. J. Mandle, A. Mertelj, Phys. Chem. Chem. Phys \textbf{21}, 18769 (2019).

\bibitem{NH} H. Nishikawa, K. Shiroshita, H. Higuchi, \textit{et. al.}  Adv. Mater. \textbf{29}(43), 1702354 (2017).
  
\bibitem{MA}  A. Mertelj, L. Cmok, Sebasti\'{a}n, \textit{et al.},  Phys. Rev. X \textbf{8}, 041025 (2018).
  
\bibitem{AM} A. Manabe, M. Bremer and M. Kraska, Liq. Cryst. \textbf{48}(8), 1079 (2021).


\bibitem{XC} X. Chen, E. Korblova, D. Dong, \textit {et al.}, Proc Nat Acad Sci. \textbf{117}, 14021 (2020).

\bibitem{NS} N. Sebasti\'{a}n, L. Cmok, R.J. Mandle, M. R. de la Fuente, I. D. Olenik, M.\u{C}opi\u{c} and A.Mertelj, Phys. Rev. Lett. \textbf{124}, 037801 (2020). 

\bibitem{REV}N. Sebasti\'{a}n, M.\u{C}opi\u{c} and A.Mertelj, Phys. Rev. E \textbf{106}, 021001 (2022).

\bibitem{PR} P. Rudquist, Sci. Rep. \textbf{11}, 24411 (2021).

\bibitem{SB} S. Brown, E. Cruickshank, J. M. D. Storey, C. T. Imrie, D. Pociecha, M. Majewska, A. Makal, E. Gorecka  Chem. Phys. Chem. \textbf{22}, 2506 (2021).

\bibitem{HN1} H. Nishikaya and F. Araoka, Adv. Mater, 2101305 (2021).

\bibitem{OD} O. D. Lavrentovich, Proc Nat Acad Sci.  \textbf{117}, 14629 (2020).

\bibitem{CLF} C. L. Folcia, J. Ortega, R. Vidal, T. Sierra and J. Etxebarria, Liq. Cryst. \textbf{49}(6), 899 (2022).

\bibitem{JLI} J. Li \textit{et al.},  Sci. Adv. \textbf{7}, eabf5047 (2021). 

\bibitem{XZ} X. Zhaoa, J. Zhoua, J. Lia, J. Kougoa, Z. Wana, M. Huanga, S. Aya,  Proc. Natl. Acd. Sci. \textbf{118}, e2111101118 (2021).

\bibitem{FC} F. Caimi, G. Nava, R. Barboza, N. A. Clark,
E. Korblova, D. M. Walba, T. Bellini and L. Lucchetti, Soft Matter \textbf{17}, 8130 (2021).

\bibitem{PK} M. Praveen Kumar, J. Karcz, P. Kula, S. Karmakar and Surajit Dhara, arXiv. https://doi.org/10.48550/arXiv.2209.11154. 

\bibitem{BI} T. C. Oakberg, Proc. SPIE  {\bf 3121}, 19 (1997).

\bibitem{ACC} T. C. Oakberg, Proc. SPIE  {\bf 2873}, 17 (1996).

\bibitem{Flexo}I. Dozov, Ph. Martinot-Lagarde and G. Durand, J. Physique  {\bf 43}, L365 (1982).

\bibitem{JT} J. Thoen, E. Korblova, D. M. Walba, N. A. Clark and C. Glorieux, Liq. Cryst. {\bf 49}, 780 (2022).

\bibitem{IH}I. Heller, Prog. Solid State Chem. {\bf 10}, 103 (1975).

\bibitem{MS} M. \v{S}karabot, K. Koc\v{e}var, R. Blinc, G. Heppke, and I. Musevic, Phys. Rev. E {\bf 59}, R1323 (1999).

\bibitem{JF} J. Fernsler, D. Wicks, D. Staines, A. Havens, and N. Paszek, Liq. Cryst. {\bf 39}, 1204 (2012).

 \bibitem{KCL} K. C. Lim and J. T. Ho, Phys. Rev. Lett.,  {\bf 40}, 1576 (1978). 
 

\bibitem{SER} S. Erkan, M. Cetinkaya, S. Yildiz, and H. \"{O}zbek, Phys. Rev. E {\bf 86}, 041705 (2012).

 \bibitem{DP} D. Pociecha, C. A. Crawford, D. A. Paterson John M. D. Storey, C. T. Imrie, N. Vaupoti\v{c}, and E. Gorecka, Phys. Rev. E {\bf 98}, 052706 (2018).


\bibitem{FL1} P. R. Maheswara Murthy, V. A. Raghunathan and N. V. Madhusudana,  Liq. Cryst. {\bf 14}, 483 (1993). 

\bibitem{FL2}  I. Dozov, P. Martinot-Lagarde, and G. Durand, J. Phys. (France) Lett. {\bf 44}, L817 (1983). 

\bibitem{CV} C. V. Brown, L. Parry-Jones, S. J. Elston, and S. J. Wilkins, Mol. Cryst. Liq. Cryst. {\bf 410}, 417 (2004).

\bibitem{HJC} H. J. Coles, M. J. Clarke, S. M. Morris, B. J. Broughton, and A. E. Blatch, J. Appl. Phys. {\bf 99}, 034104 (2006).



\end {thebibliography}
\end{document}